\documentclass[%
 reprint,
superscriptaddress,
 amsmath,amssymb,
 aps,
]{revtex4-2}

\usepackage{color}
\usepackage[colorlinks=true, allcolors=blue]{hyperref}
\usepackage{graphicx}
\usepackage{dcolumn}
\usepackage{bm}
\usepackage{hyperref}
\usepackage{soul}

\begin{document}


\title{Electronic and optical properties of crystalline nitrogen versus black phosphorus: \\ A comparative first-principles study}

\author{Alexander N. Rudenko}
\email{a.rudenko@science.ru.nl}
\affiliation{Institute for Molecules and Materials, Radboud University, 6525AJ Nijmegen, The Netherlands}

\author{Swagata Acharya}
\affiliation{Institute for Molecules and Materials, Radboud University, 6525AJ Nijmegen, The Netherlands}

 \author{Ferenc Tasn\'adi}
\affiliation{\mbox{Department of Physics, Chemistry, and Biology (IFM),
Link\"{o}ping University, SE-581 83, Link\"oping, Sweden}}

\author{Dimitar Pashov}
\affiliation{Theory and Simulation of Condensed Matter, King's College London, The Strand, WC2R 2LS London, UK}

\author{Alena V. Ponomareva}
\affiliation{Materials Modeling and Development Laboratory, NUST ``MISIS'', 119049 Moscow, Russia.}

\author{Mark van Schilfgaarde}
\affiliation{Theory and Simulation of Condensed Matter, King's College London, The Strand, WC2R 2LS London, UK}
\affiliation{National Renewable Energy Laboratory, 15013 Denver W Pkwy, Golden, Colorado 80401, USA}

\author{Igor A. Abrikosov}
\affiliation{\mbox{Department of Physics, Chemistry, and Biology (IFM),
Link\"{o}ping University, SE-581 83, Link\"oping, Sweden}}
 
\author{Mikhail I. Katsnelson}
\affiliation{Institute for Molecules and Materials, Radboud University, 6525AJ Nijmegen, The Netherlands}

\date{\today}

\begin{abstract}
Crystalline black nitrogen (BN) is an allotrope of nitrogen with the black phosphorus (BP) structure recently synthesized at high pressure by two independent research groups [Ji \emph{et al}., Sci. Adv. {\bf 6}, eaba9206 (2020); Laniel \emph{et al}., Phys. Rev. Lett. {\bf 124}, 216001 (2020)]. Here, we present a systematic study of the electronic and optical properties of BN focusing on its comparison with BP. To this end, we use the state-of-the-art quasiparticle self-consistent $GW$ approach with vertex corrections in both the electronic and optical channels. Despite many similarities, the properties of BN are found to be considerably different. Unlike BP, BN exhibits a larger optical gap (2.5 vs 0.26 eV), making BN transparent in the visible spectral region with a highly anisotropic optical response. This difference can be primarily attributed to a considerably reduced dielectric screening in BN, leading to enhancement of the effective Coulomb interaction. Despite relatively strong Coulomb interaction, exciton formation is largely suppressed in both materials. Our analysis of the elastic properties shows exceptionally high stiffness of BN, comparable to that of diamond.
\end{abstract}

\maketitle


\section{\label{sec1}Introduction}
Nitrogen is one of the most abundant elements on Earth, occurring in a gaseous or liquid form. Solid nitrogen is by far less common and is mostly known in the form of molecular N$_2$ crystals, which are, for instance, a main component of Pluto \cite{Sternaad1815}.
Other forms of nitrogen include polymeric forms \cite{PhysRevLett.102.065501,Laniel2019} and crystals with the highly unusual cubic gauche structure  \cite{PhysRevB.46.14419, Eremets2004} stabilized at high pressure. Another allotrope of nitrogen, so-called black nitrogen (BN), was recently synthesized at high pressure by two independent research groups \cite{Jieaba9206,Laniel2020}.
BN adopts the orthorhombic (A17) crystal structure, identical to that of black phosphorus (BP).

Exotic forms of matter often demonstrate unusual properties which attract interest from the research community. In high-pressure physics, the most prominent example is metallic hydrogen, an elusive material expected to exhibit a variety of remarkable properties, including high-temperature superconductivity \cite{PhysRevLett.21.1748} and the ultimate speed of sound \cite{doi:10.1126/sciadv.abc8662}.
Understanding the properties of previously unknown phases of elemental compounds is important for fundamental science as it complements our knowledge of trends running through the periodic table. In addition, studies of solid nitrogen are relevant in the context of high-energy-density materials. 

Among the group-V materials, BP is a well-known and thoroughly studied material, which was rediscovered several years ago from the perspective of a two-dimensional material \cite{LXing2015,FXia2019,YiXu2019}. Unlike BP, the properties of BN are largely unexplored. Apart from basic spectroscopic characteristics \cite{Jieaba9206,Laniel2020}, information about their microscopic origin and the physical mechanisms behind the observable properties remain unclear. In this work, we perform a systematic first-principles analysis of the electronic, optical, and vibrational properties of BN under the experimental pressure conditions. We start from density functional calculations and use the quasiparticle self-consistent $GW$ method with vertex corrections to accurately describe the optical response. We focus on the comparison with BP and underline mechanisms responsible for the difference between the properties of BN and BP. 

The rest of this paper is organized as follows. In Sec.~\ref{sec2}, we briefly describe theoretical methods and provide computational details. In Secs.~\ref{sec3a} and \ref{sec3b}, we present our results on the electronic structure and optical properties of BN and BP, which are calculated at different levels of theory. We then analyze the Coulomb interaction and screening in these two materials (Sec.~\ref{sec3c}). Sec.~\ref{sec3d} is devoted to a comparative analysis of the vibrational characteristics and elastic properties. In Section~\ref{sec4}, we briefly summarize our results and conclude the paper.

\section{\label{sec2}Computational details}
\emph{Electronic and optical properties.}
We study the electronic structure and optical properties of BN/BP at three different levels of theory: density functional theory (DFT) within the local density approximation (LDA), the quasiparticle self-consistent
	\emph{GW} approximation (QS\emph{GW})~\cite{kotani,pashov}, and an extension of QS\emph{GW} in which the random phase approximation (RPA)
	to the polarizability is extended by adding ladder diagrams (QS$G\widehat{W}$)~\cite{bseoptics,brian}.
		The optical properties are calculated by incorporating the electron-hole two-particle correlations within a self-consistent ladder-Bethe-Salpeter-equation (BSE) implementation~\cite{bseoptics,brian} with the Tamm-Dancoff
	approximation~\cite{tamm,dancoff}.
	For BP (BN), DFT calculations and energy band calculations with the static quasiparticle QS\emph{GW} and QS$G\widehat{W}$ self-energy
	$\Sigma^{0}(k)$ were performed on a 24$\times$24$\times$12 (18$\times$14$\times$6) \emph{\bf k} mesh, while the dynamical self-energy
	$\Sigma({\bf k})$ was constructed using an 8$\times$8$\times$4  (9$\times$7$\times$3) \emph{\bf k} mesh and $\Sigma^{0}$({\bf k}) was extracted from it.  For each
	iteration in the QS\emph{GW} and QS$G\widehat{W}$ self-consistency cycles, the charge density was made self-consistent.  The QS\emph{GW} and QS$G\widehat{W}$ cycles
	were iterated until the rms change in $\Sigma^{0}$ reached 10$^{-5}$\,Ry.  Thus, the calculation was self-consistent in
	both $\Sigma^{0}({\bf k})$ and the density. We observe that
	for BP the QS$G\widehat{W}$ band gap stops changing once 4 valence and 2 conduction bands are included in the
	two-particle Hamiltonian, while for BN, the convergence was achieved with 12 valence and 12 conduction bands. We use a \emph{\bf k} mesh of 24$\times$24$\times$12 and 9$\times$14$\times$5 for computation of the real and imaginary parts of the dielectric response functions for BP and BN, respectively. In both cases, an energy-dependent optical broadening parameter that varies from 10 meV at $\omega$=0 to 1 eV at $\omega$=27.2 eV is used. The {\bf k}-point convergence of the dielectric function is shown in the Appendix (Fig.~\ref{conv-test}). The electronic structure calculations have been performed on a conventional orthorhombic unit cell containing eight atoms. Experimental lattice parameters were used in all cases.

\emph{Coulomb interactions.}
Coulomb interactions were calculated in the Wannier function (WF) basis using the procedure implemented in {\sc vasp} \cite{Kresse1,Kresse2,vasp-gw1,vasp-gw2,KaltakcRPA}. For this purpose, we carried out calculations within the projector augmented wave formalism (PAW) \cite{paw1,paw2} using the generalized gradient approximation (GGA) functional in the Perdew-Burke-Ernzerhof parametrization \cite{pbe}.  A 400 (250) eV energy cutoff for the plane-waves and a convergence threshold of $10^{-8}$ eV were used for BN (BP). The calculations for both compounds were performed using a primitive cell containing four atoms.
The Brillouin zone was sampled by a 6$\times$6$\times$8 ${\bf k}$-point mesh. The WFs were calculated considering four WFs per atom ($sp$ basis) using the scheme of maximal localization \cite{mlwf1,mlwf2} using the {\sc wannier90} package \cite{wannier90}. The screening was considered at the full RPA level \cite{CoulombU} taking \emph{all} possible empty states into account within the given plane-wave basis. 
Specifically, we used 375 (696) bands in total, respectively, for BN (BP).
The effective on-site and intersite Coulomb interaction was calculated by averaging the orbital components of the Coulomb matrix as $U_{\mathrm{eff}}({\bf R}_{ij}) = 1/16\sum_{mn}U_{mn}({\bf R}_{ij})$, where $i$ and $j$ are the atomic indices and $m$ and $n$ are the orbital indices.

\begin{figure}[t]
\includegraphics[scale=0.33]{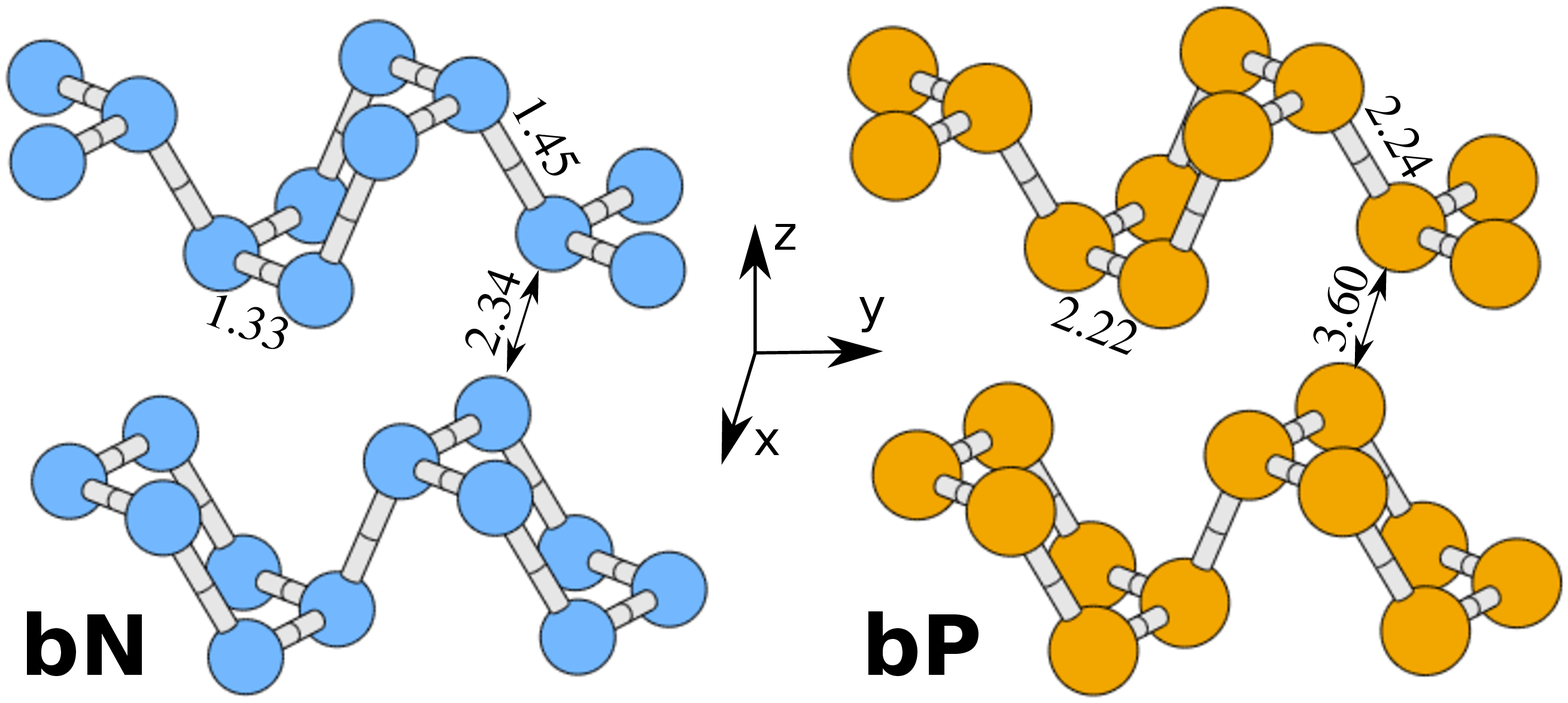}

\vspace{0.2cm}

\includegraphics[scale=1.8]{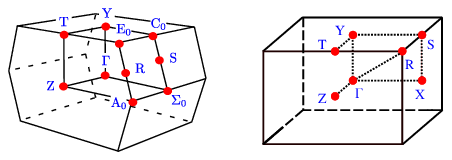}
\caption{\label{structure} Top: Schematic crystal structure of BN (left) and BP (right). The numbers correspond to the bond lengths (in \AA). Bottom: Brillouin zone with high-symmetry points for primitive (left) and conventional (right) unit cells of BN/BP containing four and eight atoms, respectively.}
\end{figure}

\emph{Phonon excitations and elastic properties.}
The vibrational and elastic properties were calculated with DFT as implemented
in {\sc vasp} \cite{Kresse1,Kresse2} using the
PAW formalism \cite{paw1,paw2}.
The exchange-correlation energy was approximated by the Perdew-Burke-Ernzerhof GGA functional \cite{pbe}. A 540\,eV energy cutoff and a $5\times 5\times 5$ ${\bf k}$-point sampling of the Brillouin zone were used with $4\times 4\times 4$ supercells, which is sufficient to obtain converged phonon dispersions. The phonon calculations were performed using the finite-displacement method implemented in {\sc phonopy} \cite{togo_first_2015}. 
The pressures were derived from the diagonal elements of the stress tensor calculated for the
supercells with optimized atomic positions. The Raman-active modes were derived by factor group analysis
of the vibrations \cite{deangelis_factor_1972} at the $\Gamma$ point. The
elastic constants $C_{ij}$ were calculated using enthalpy-strain relationships applying $\pm1\%$ and $\pm2\%$ strain
to the primitive cell containing four atoms. In these calculations, we used a $12\times 12\times 12$ {\bf k}-point mesh to sample the Brillouin zone. 
The sound velocities were derived by solving the Christoffel equation based on the
elastic constants.
The anisotropic Young's modulus $E({\bf n})$ was calculated as defined in \cite{ting_anisotropic_2006} 
using a $201\times51$ mesh of the polar and azimuth angles defining the directional normal vector ${\bf n}$.

\begin{figure}[t]
\includegraphics[scale=0.73]{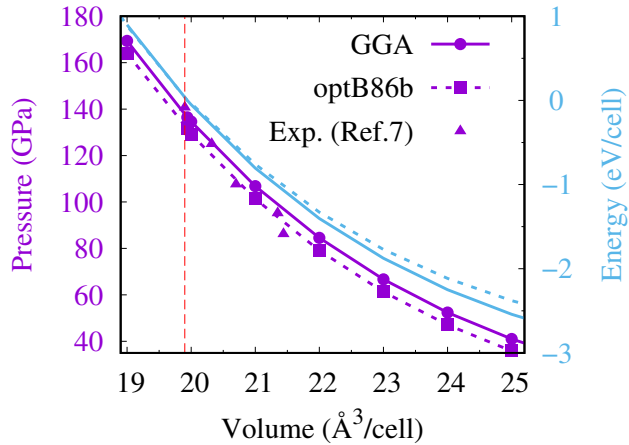}
\caption{\label{P-E-V} Pressure and total energy shown as a function of the primitive cell volume calculated for BN with the GGA and optB86b functionals. Experimental pressures from Ref.~\cite{Laniel2020} are shown for comparison. The vertical line corresponds to the experimental volume ($V_\mathrm{exp}\approx20$ \AA$^3$) used in this work. Zero total energy corresponds to the energy calculated at $V_\mathrm{exp}$.
}
\end{figure}

\emph{Crystal structure}.
The crystal structures of BN and BP are schematically shown in Fig.~\ref{structure} along with the Brillouin zones of the primitive and conventional unit cells. The structure consists of puckered layers stacked along the $z$ direction and corresponds to the A17 phase with space group $Cmca$ (No.~64). For BP, this structure is thermodynamically stable up to $\sim$5 GPa, at which it transforms into a layered rhombohedral (A7) phase \cite{Leoni2012,Scelta2017}. In contrast, BN is unstable under ambient conditions but can be stabilized at a high pressure of 140--150 GPa \cite{Laniel2020,Jieaba9206}. 
Here, we use these conditions and the corresponding experimental lattice parameters to study the properties of BN.
With the experimental lattice parameters used in this work, the hydrostatic stress calculated using GGA corresponds to 138 (2.4) GPa for BN (BP).

\begin{table}[b]
\caption{\label{tab:table1}%
Band gaps (eV) calculated for BN and BP at different levels of theory.
The BSE optical gap is determined by the energy of the deepest-lying bright exciton.}
\begin{ruledtabular}
\begin{tabular}{lccccc}
\textrm{}&
\textrm{DFT}&
\textrm{QS$GW$\footnote{Single-particle gap.}}&
\textrm{QS$G\widehat{W}$}$^{\mathrm{a}}$&
\textrm{RPA@QS$G\widehat{W}$}$^{\mathrm{b}}$&
\textrm{BSE@QS$G\widehat{W}$\footnote{Optical gap.}}\\
\colrule
BN & metal & 1.88 & 1.61 & 3.0 & 2.50 \\
BP & metal & 0.30 & 0.28 & 0.28 & 0.26 \\
\end{tabular}
\end{ruledtabular}
\end{table}

It should be noted that, experimentally, it was possible to decompress BN down to 48 and 86 GPa after the synthesis (Refs.~\cite{Jieaba9206,Laniel2020}, respectively). Figure \ref{P-E-V} shows pressure and total energy as a function of the primitive cell volume calculated for BN with (GGA) and without (optB86b \cite{Klimes}) van der Waals interactions taken into account. Both exchange-correlation functionals yield essentially similar behavior, indicating that the van der Waals interactions do not play a significant role in the thermodynamics of BN. The pressure vs volume dependence is in agreement with the experimental data from Ref.~\onlinecite{Laniel2020}.
For the volumes considered, pressure behaves monotonically, spanning a wide range of values, suggesting stability of the A17 structural phase upon decompression down to at least 40 GPa. The dynamical stability of this phase is also confirmed by the absence of imaginary modes in the phonon spectrum of BN above 22 GPa.

\section{\label{sec3}Results}

\begin{figure}[t]
\includegraphics[scale=0.25]{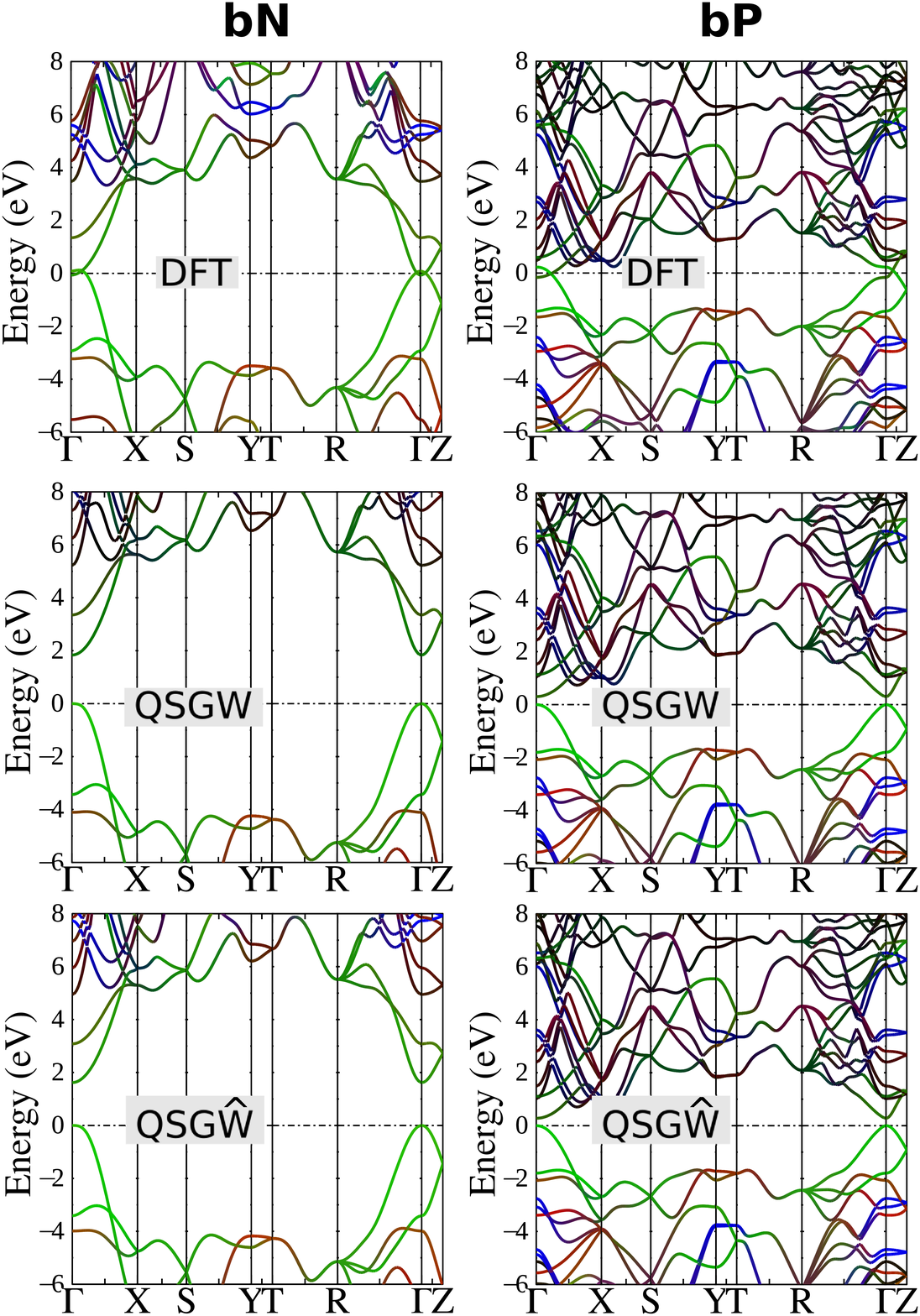}
\caption{\label{bands} Single-particle band structure of BN (left) and BP (right) calculated using the DFT (LDA), QS$GW$, and QS$G\widehat{W}$ approaches (from top to bottom) along high-symmetry directions of the conventional Brillouin zone (Fig.~\ref{structure}). The colors correspond to the contributions of $p_x$ (blue), $p_y$ (red), and $p_z$ (green) orbitals.
}
\end{figure}

\subsection{\label{sec3a}Electronic structure}
Figure~\ref{bands} shows single-particle band structure of BN and BP calculated using DFT (LDA), QS$GW$, and QS$G\widehat{W}$ methods. Table \ref{tab:table1} summarizes the calculated band gaps. At the DFT level, both BN and BP are metals, showing overlap between the valence and conduction bands, with negative (inverted) gaps of $-$0.15 and $-$0.4 eV, respectively.  The single-shot $GW$ based on LDA produces a positive gap of $\sim$0.1 eV in BP and 0.8 eV in BN. Both QS$GW$ and QS$G\widehat{W}$ result in the formation of a direct gap at the $\Gamma$ point. For BP, a gap of $\sim$0.3 eV is found,
in agreement with previous $GW$ calculations \cite{Rudenko2015}, as well as with experimental measurements \cite{Keyes1953,Kiraly2017,Carr__2021}. For BN, the gap turns out to be significantly larger, reaching 1.6 eV at QS$G\widehat{W}$, which suggests that screening in BN is smaller than in BP. This can be attributed to the suppression of the Coulomb screening in BN, as will be shown in Sec.~\ref{sec3c}.

\begin{figure}[t]
\includegraphics[scale=0.55]{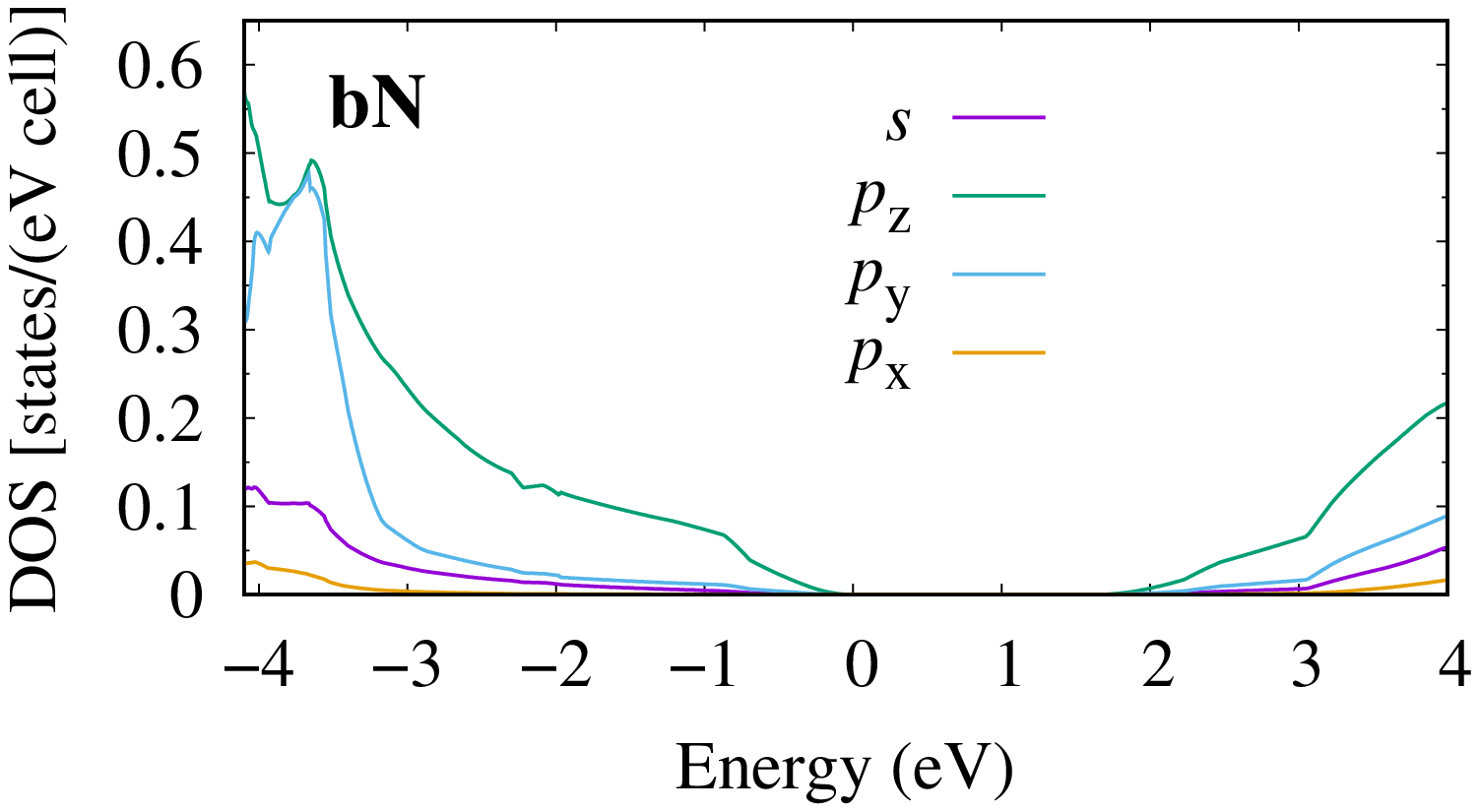}
\includegraphics[scale=0.55]{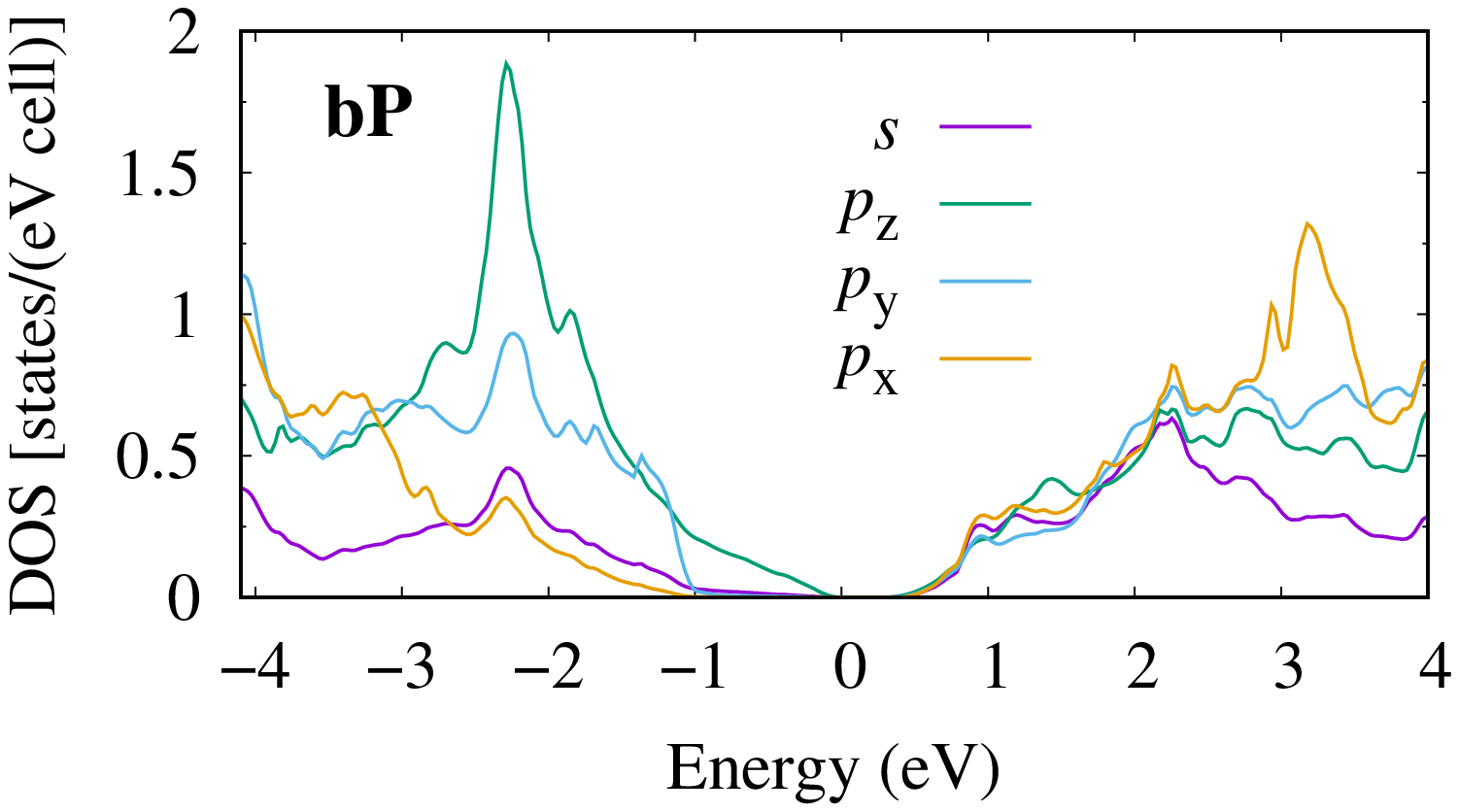}
\caption{\label{dos} Density of states projected onto different orbital states calculated for BN (top) and BP (bottom) using the QS$G\widehat{W}$ approach.}
\end{figure}

Figure~\ref{dos} shows the density of states (DOS) projected onto different orbital contributions in BN and BP calculated using QS$G\widehat{W}$. Unlike BP, where the conduction band constitutes a mixture of the $s$ and $p$ states, the conduction band of BN is mainly composed of the $p_z$ states. In contrast, the primary contribution to the valence band edge results from the $p_z$ states in both materials. Overall, $sp$ electronic states of BN are characterized by a smaller density, spanning a larger energy range compared to BP.

\begin{table}[b]
\caption{\label{tab:table2}%
Electron and hole effective masses (in units of free-electron mass) calculated for BN and BP along three crystallographic directions using the QS$GW$ and QS$G\widehat{W}$ approximations.
}
\begin{ruledtabular}
\begin{tabular}{ccccccc} 
 & \multicolumn{3}{c}{BN} & \multicolumn{3}{c}{BP} \\
 \cline{2-4} \cline{5-7}
\textrm{}&
\textrm{$m_x$ }&
\textrm{$m_y$ }&
\textrm{$m_z$ }&
\textrm{$m_x$ }&
\textrm{$m_y$ }&
\textrm{$m_z$ } \\
\colrule
Electrons\footnote[1]{QS$GW$ approximation.} & 1.19 & 0.14 & 0.52 & 1.11 & 0.09 & 0.12 \\
Electrons\footnote[2]{QS$G${$\widehat{W}$}\text{ approximation.}} & 1.25 & 0.14 & 0.54 & 1.11 & 0.09 & 0.12 \\
\hline
Holes\footnotemark[1] & 6.11  & 0.23 &  0.51  & 0.67 &  0.08 &  0.28 \\
Holes\footnotemark[2] & 12.14 & 0.23 &  0.52  & 0.67 &  0.08 &  0.28
\end{tabular}
\end{ruledtabular}
\end{table}

Table \ref{tab:table2} summarizes the effective masses calculated for BN and BP using the QS$GW$ and QS$G\widehat{W}$ methods. The electron effective masses in the $xy$ direction are similar for BN and BP, being highly anisotropic with the ratio $m_x/m_y \sim 10$. In the direction perpendicular to the layer stacking ($z$) the effective mass of BN ($m_z\approx0.5m_e$) is significantly larger than in BP ($m_z\approx0.1m_e$). 
These results are consistent between the QS$GW$ and QS$G\widehat{W}$ methods. 
Larger $m_z$ implies that the states are less dispersed along the $k_z$ direction, indicating a weaker overlap between the wave functions of the interacting layers. This overlap is not favorable energetically as the corresponding interlayer hopping integrals are positive \cite{Rudenko2015}. This means that the band contribution to the interlayer binding is repulsive, and the repulsion is weaker in BN.

The situation with the hole effective masses is qualitatively similar between the materials with one exception. Namely, the hole effective masses calculated along the zigzag ($x$) direction in BN are anomalously high, resulting in $6.1m_e$ and $12.1m_e$ within QS$GW$ and QS$G\widehat{W}$, respectively, which is an order of magnitude larger compared to BP. We note that this anomaly is not related to any topological peculiarities of the Fermi surface, as indicated by the absence of the Van Hove singularities in the calculated DOS (Fig.~\ref{dos}). Instead, this mass enhancement can be solely attributed to the correlation effects. Unlike all other cases, $m_x$ for holes is considerably different within QS$GW$ and QS$G\widehat{W}$, which indicates an important role of the vertex corrections. We do not exclude that higher-order diagrams, not considered here, might also be important for a correct determination of $m_x$ in BN.

\subsection{\label{sec3b}Optical response}

\begin{figure*}[tbp]
\includegraphics[scale=0.55]{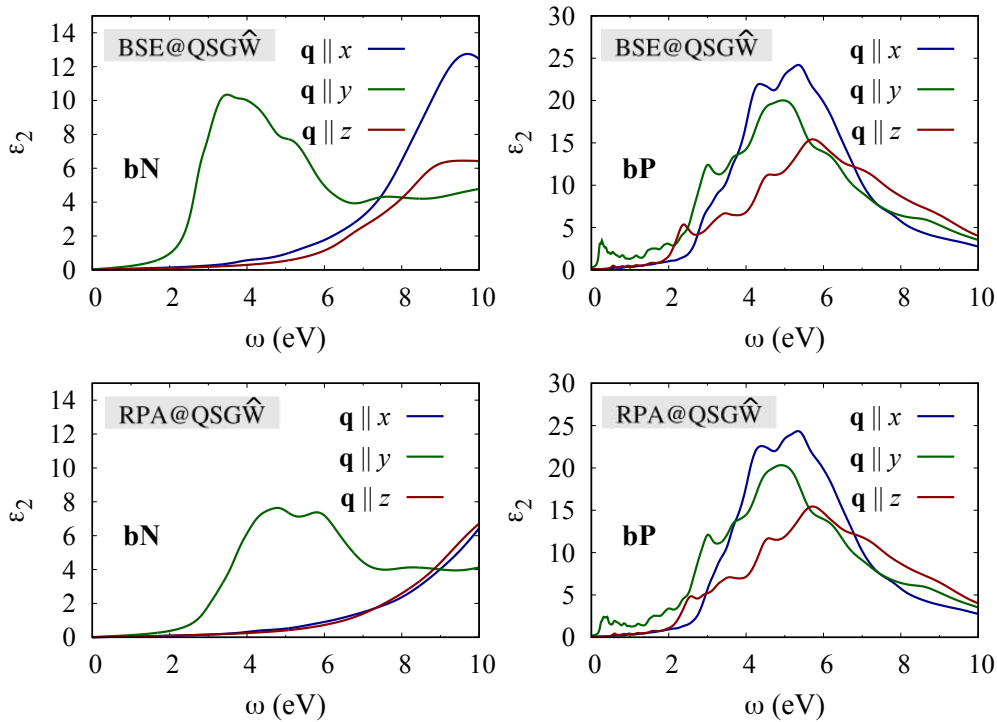}
\caption{\label{optics} Imaginary part of the diagonal components of the dielectric tensor calculated as a function of frequency using RPA@QS$G\widehat{W}$ (top) and BSE@QS$G\widehat{W}$ (bottom) for BN (left) and BP (right). }

\end{figure*}

In our calculations, we use a rectangular unit cell; thus, the crystal axes $x$, $y$, and $z$ correspond to the principal axes. In this situation the dielectric tensor $\varepsilon^{\alpha \beta}$ is diagonal in the absence of magnetic fields.
Figure~\ref{optics} shows the imaginary parts of the corresponding frequency-dependent diagonal components $\varepsilon^{\alpha \alpha}_{2}$ ($\alpha=x,y,z$) calculated at the RPA@QS$G\widehat{W}$ and BSE@QS$G\widehat{W}$ levels for BN and BP. In the relevant spectral region, the strongest absorption is along the $y$ direction (armchair) for both BN and BP. This can be attributed to low effective masses in the corresponding direction (see Table \ref{tab:table2}), which is also reflected in the DOS (Fig.~\ref{dos}). On the other hand, large effective masses along $x$ and $z$ suppress the dipole transition matrix elements, which determine the optical absorption. It should be noted that the corresponding matrix elements are not symmetry forbidden at the onset of $\varepsilon_2(\omega)$.

For BP, the optical gap obtained within RPA coincides with the single-particle band gap (see Table \ref{tab:table1}), as expected. The vertex corrections applied at the BSE level slightly reduce the gap, yielding 0.26 eV of optical gap (20 meV exciton binding energy), which is consistent with previous theoretical studies, as well as with recent experiments \cite{Carr__2021}. No new features appear in the spectrum after the inclusion of the vertex corrections, demonstrating that their effect is not significant, in agreement with earlier findings \cite{Tran2014}.

\begin{figure}[b]
\mbox{\includegraphics[scale=0.36]{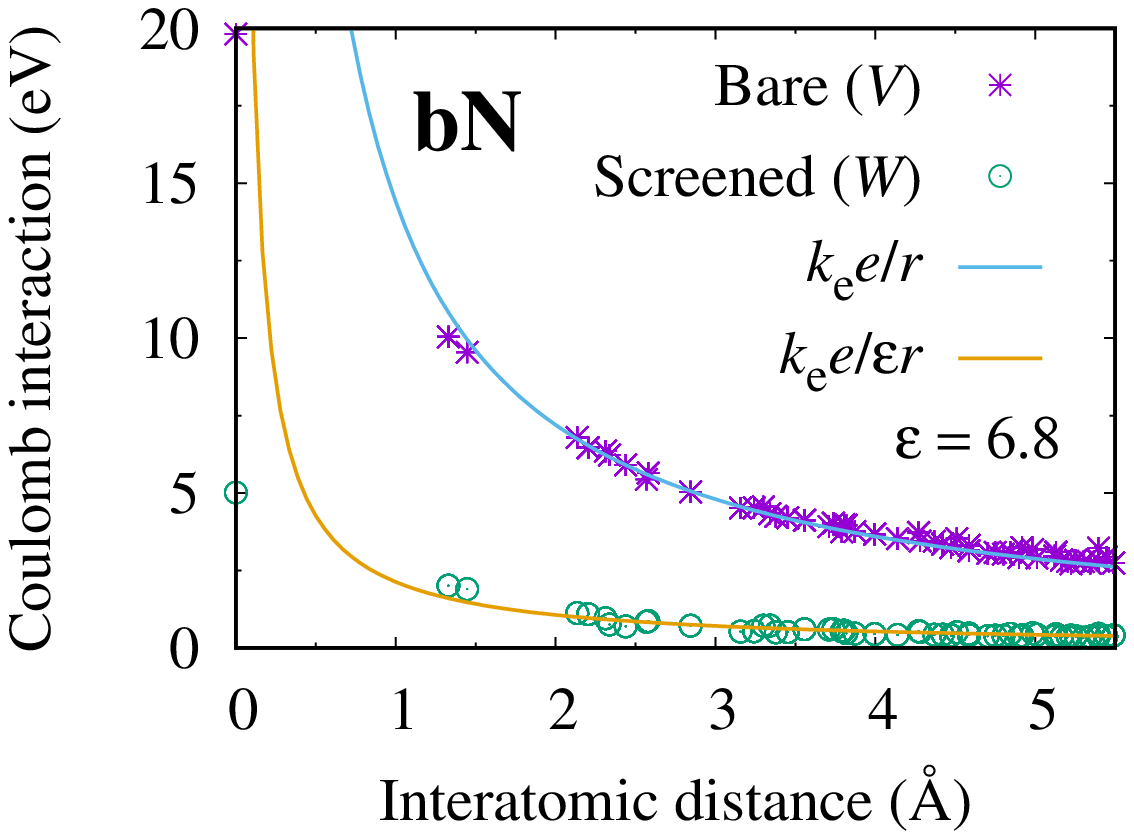}
\includegraphics[scale=0.35]{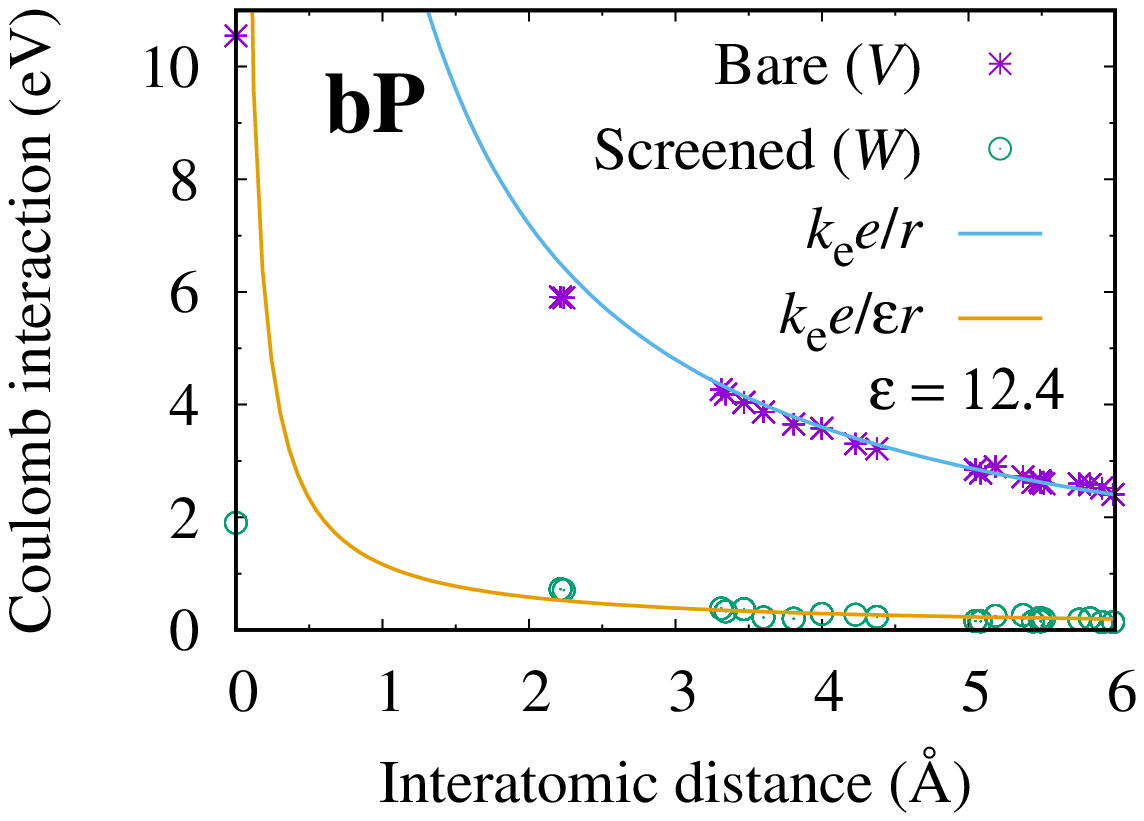}
}
\caption{\label{coulomb} Effective Coulomb interaction as a function of the interatomic distance in BN and BP. Violet and green symbols correspond to the bare ($V$) and screened ($W$) interactions, respectively. Lines correspond to the classical $V(r)=k_ee/r$ and screened classical $W(r)=V(r)/\varepsilon$ Coulomb interaction, where $\varepsilon$ is the effective screening constant.}
\end{figure}

The situation with BN is less trivial. Already, the RPA calculations show that the optical gap (3.0 eV) is dramatically different from the single-particle gap (1.6 eV). This can be ascribed to the orbital composition of the valence and conduction bands in BN (see the projected DOS in Fig.~\ref{dos}). In contrast to BP, the $s$ states do not contribute to the valence and conduction band edges, which are predominantly composed of the $p_z$ states. This means that the dipole transitions between the band edges are symmetry forbidden, resulting in a vanishing optical absorption at the single-particle gap energies. At higher energies, as different angular momenta mix, the dipole transitions become allowed, ensuring finite absorption at 3.0 eV. The presence of vertex corrections further reduces this value by $\sim$0.5 eV, yielding an optical gap of 2.5 eV. 
\begin{table}[b]
\caption{\label{tab:table3}%
Effective on-site and intersite Coulomb interaction for BN and BP without ($V$) and with ($W$) screening effects considered at the full RPA level. 1NN and 2NN denote nearest and next-nearest neighbors, respectively. $d$ is the distance between the corresponding atoms.
}
\begin{ruledtabular}
\begin{tabular}{ccccccc}
 & \multicolumn{3}{c}{BN} & \multicolumn{3}{c}{BP} \\ \cline{2-4} \cline{5-7}
\textrm{}&
\textrm{On site}&
\textrm{1NN}&
\textrm{2NN}&
\textrm{On site}&
\textrm{1NN}&
\textrm{2NN} \\
\colrule
$d$ (\AA) & -- & 1.33 & 1.45  & -- & 2.22 & 2.24 \\
$V$ (eV) & 19.8 & 10.0 & 9.5 & 10.5 & 5.9  &  5.9 \\
$W$ (eV) & 5.0 & 2.0 & 1.9  & 1.9 & 0.7  & 0.7 \\
\end{tabular}
\end{ruledtabular}
\end{table}
Therefore, our BSE@QS$G\widehat{W}$ calculations
demonstrate that pristine BN can absorb the blue part of the visible spectrum. Additionally, we find four dark $e$-h eigenvalues at lower energies (1.35, 1.87, and 1.92 eV and two degenerate eigenvalues at 2 eV) compared to the optical gap edge. Although
these dark excitons are not important for optical absorption, they might be important for exciton dynamics and photoluminescence. 
Nevertheless, for both BP and BN, mostly, the valence band maximum and conduction band bottom contribute to the low-energy optical absorption. The excitons in these materials are Wannier-Mott type, in contrast to the Frenkel excitons in compounds like Cr$X_{3}$~\cite{swagcrx}, where all Cr $d$ bands participate in the low-energy optical absorption.

\begin{table}[t]
\caption{\label{tab:table4}%
Diagonal components of the ion-clamped static dielectric tensor $\varepsilon^{\infty}_{ii}$ ($ii=xx,yy,zz$) derived from the BSE@QS$G\widehat{W}$ calculations. 
}
\begin{ruledtabular}
\begin{tabular}{cccc}
\textrm{}&
$xx$  &
$yy$  &
$zz$ \\
\colrule
BN & 7.1 & 9.5 & 6.0 \\
BP & 12.9 & 16.1 & 10.4    
\end{tabular}
\end{ruledtabular}
\end{table}

\subsection{\label{sec3c}Coulomb interaction and screening}
To explain the huge difference between the band gaps in BN and BP, we analyze the Coulomb interactions in both materials. 
Figure~\ref{coulomb} shows the effective Coulomb interaction calculated for BN and BP, where both bare ($V$) and fully screened ($W$) values are plotted as a function of the interatomic distance. The on-site interactions and intersite interactions up to the second-nearest neighbor are also summarized in Table \ref{tab:table3}. The bare on-site interaction in BN (19.9 eV) is larger by a factor of $\sim$1.9 compared to that in BP (10.5 eV). This can be understood in terms of the difference in orbital localization in these two materials. Indeed, an average quadratic spread of the corresponding $sp$ Wannier orbitals obtained within the maximum-localization procedure is found to be 0.5 and 2.7 \AA$^2$ for BN and BP, respectively. The ratio of the \emph{screened} on-site interactions is only slightly larger ($W^{\mathrm{BN}}/W^{\mathrm{BP}}\sim 2.6$), indicating that the local screening effects are comparable in BN and BP. The situation with the interatomic (nonlocal) interactions is different. Interestingly, already from the nearest neighbor, the bare interaction is well described by the classical Coulomb potential $V(r)=k_e e/r$ for both BN and BP, where $k_e=1/4\pi\varepsilon_0$ is the Coulomb constant. The screened intersite interaction can be reasonably well fitted by the potential $W(r)=V(r)/\varepsilon$, with $\varepsilon$ being an effective screening constant. The values of $\varepsilon$ are estimated as 6.8 and 12.4 for BN and BP, respectively, indicating that the screening in BP is considerably more efficient. At the same time, strong screening can be related to large DOS (see Fig.~\ref{dos}) in a wide region around the band gap, resulting in an enhancement of the optical spectral weight at all relevant frequencies. 
Overall, the Coulomb interactions in BN are considerably larger compared to those in BP. At the level of the perturbation theory, this means that many-body corrections  should lead to a larger self-energy, which explains the wider band gap observed in BN. We note that the obtained screened interactions might be somewhat overscreened and therefore underestimated because the calculations are based on the DFT band structure without many-body corrections.

\begin{figure}[t]
\includegraphics[scale=1.15]{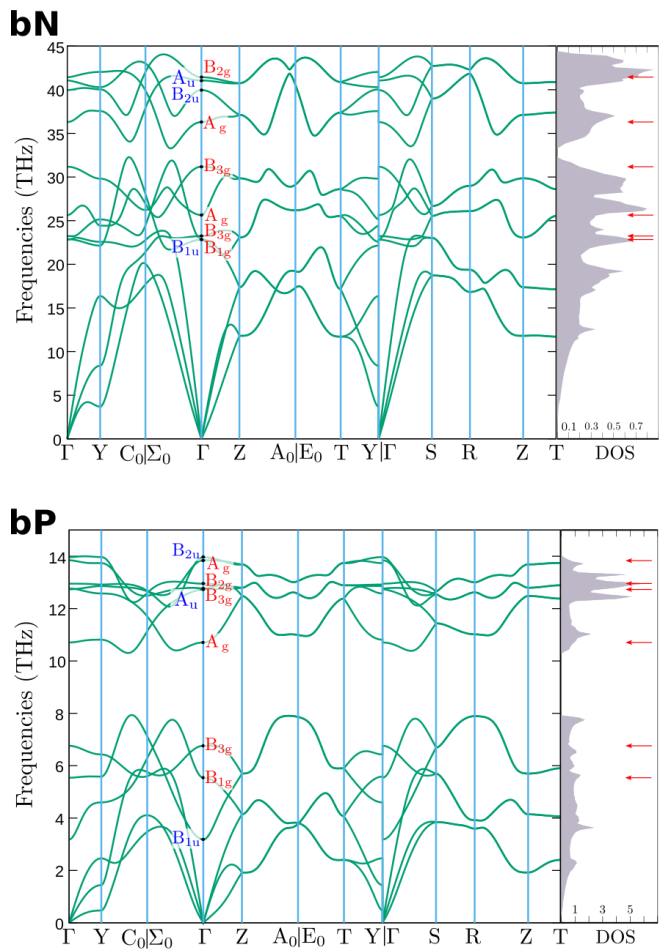}
\caption{\label{phonons} Phonon dispersion and phonon density of states calculated along the high-symmetry direction of the primitive Brillouin zone (Fig.~\ref{structure}) for BN and BP. The labels at the $\Gamma$ points indicate the symmetries of the corresponding optical modes according to irreducible representations of the $D_{2h}$ point group. Red arrows denote frequencies of the Raman-active modes.}
\end{figure}

\begin{table}[b]
\caption{Raman-active phonon modes (THz) calculated for BN and BP shown in accordance with irreducible representations of the $D_{2h}$ point group.\label{raman}}
\centering
\begin{ruledtabular}
\begin{tabular}{ccccccc}
 &  $B_{1\text{g}}$  & $B_{3\text{g}}$ & $A_{\text{g}}$ & $B_{3\text{g}}$ & $A_{\text{g}}$ &  $B_{2\text{g}}$ \\
\hline
BN & 22.9 & 23.2 & 25.6 & 31.2 & 36.3 & 41.4 \\
BP & 5.6 & 6.8 & 10.7 & 12.8 & 13.9 & 13.0 \\
\end{tabular}
\end{ruledtabular}
\end{table}

To gain more insight into the dielectric screening in BN and BP, in Table \ref{tab:table4} we provide the ion-clamped static dielectric function $\varepsilon^{\infty}_{ii}$ ($ii=xx,yy,zz$) extracted from the BSE-corrected optics and resolved in three crystallographic directions.
In agreement with the results presented in Fig.~\ref{coulomb}, we can see that the screening in BP is considerably larger than in BN in all three directions.
In addition, the screening is highly anisotropic in both materials and is highest in the armchair ($y$) direction. This observation is consistent with the fact that the optical edge is mainly determined by the $yy$ component of the dielectric function (Fig.~\ref{optics}).

\subsection{\label{sec3d}Vibrational and elastic properties}

\begin{figure}[tbp]
\includegraphics[scale=1.00]{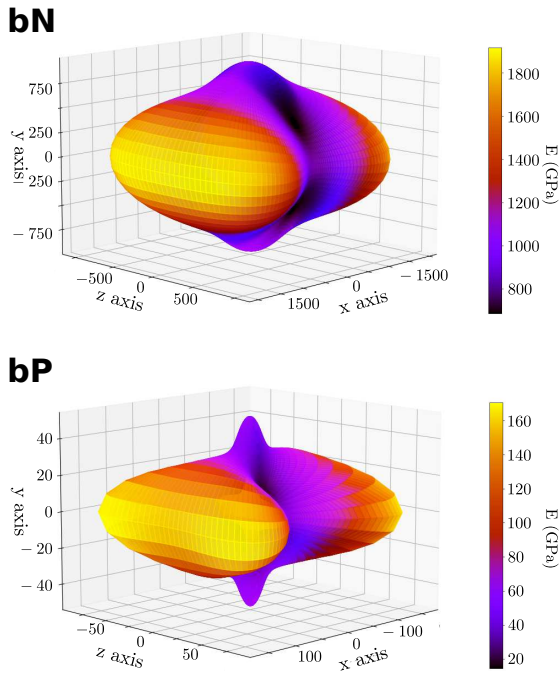}
\caption{\label{young}Orientation dependence of Young's modulus $E({\bf n})$ in BN and BP.}
\end{figure}

Figure \ref{phonons} shows the calculated phonon dispersions and DOS.
Although the dispersions show a large degree of
similarity between BN and BP, we observe much higher vibrational
frequencies in BN.
While the largest optical frequency in BP is found around 14 THz, the corresponding value reaches 44 THz in BN, i.e., more than 3 times higher. This difference cannot be explained by the mass difference between N and P atoms,  $\omega^{\mathrm{BP}}/\omega^{\mathrm{BN}} \neq \sqrt{m_N/m_P}\approx 1.5 $.
Apparently, it is attributed to stronger interatomic interactions in BN, which is exaggerated by external pressure.

In systems with $D_{2h}$ point group symmetry like BN and BP, the zone center optical phonons can be classified according to irreducible representations of the point group as follows:
\begin{equation}
\notag
    \Gamma_{D_{2h}} = 2A_g + B_{1g} + B_{2g} + 2B_{3g} + A_u + 2B_{1u} +2B_{2u}+B_{3u}.
\end{equation}
The corresponding mode symmetries are indicated in Fig.~\ref{phonons} by blue and red labels. The modes with \emph{gerade} parity (red) can be characterized as Raman active, and their frequencies are listed in Table \ref{raman}. Our results for both BN and BP agree
well with those reported earlier in Refs.~\onlinecite{appalakondaiah_effect_2012,ribeiro_raman_2018,Laniel2020}.

The low-frequency phonons shown in Fig.~\ref{phonons} are characterized by one longitudinal and two transverse branches. In order to gain more insight into the behavior at long wavelengths, we have calculated the sound velocities for different crystallographic directions and polarization; the results are summarized in Table ~\ref{velocities}. As expected, the sound velocities are highly anisotropic in both materials. It is worth noting that the anisotropy of elastic-related properties is considerably smaller in BN, which is likely related to the effect of pressure.
The highest velocities are found for phonons propagating in the zigzag ($x$) directions with the longitudinal polarization. In this case, we obtain 21.1 km/s for BN, which is larger than the sound velocity in diamond, and only 1.7 times smaller than the ultimate speed of sound \cite{doi:10.1126/sciadv.abc8662}.
In BP, the resulting values are 2--3 times smaller compared to those for BN.

\begin{table}
\caption{Sound velocities (km/s) calculated for different crystallographic directions for longitudinal (L) and two transverse (T$_1$ and T$_2$) components of the polarization in BN and BP. $\Theta_{\mathrm{D}}$ is the Debye temperature (K).
\label{velocities}}
\centering
\begin{ruledtabular}
\begin{tabular}{ccccccc}
 & \multicolumn{3}{c}{BN} & \multicolumn{3}{c}{BP} \\
 \cline{2-4} \cline{5-7}
 &  L  &  T$_1$ & T$_2$ &  L  & T$_1$ &  T$_2$ \\
\hline
$v_x$ & 21.1 & 9.3 & 12.6 & 8.4 & 2.5 & 4.8 \\
$v_y$ & 14.9 & 9.3 & 7.4 & 4.4 & 2.5 & 1.3 \\
$v_z$ & 13.7 & 12.6 & 7.4 & 4.1 & 4.8 & 1.3 \\
\hline
$\Theta_\mathrm{D}$ & \multicolumn{3}{c}{1840} & \multicolumn{3}{c}{314}
\end{tabular}
\end{ruledtabular}
\end{table}

In Fig.~\ref{young}, we show a comparison between the orientation-dependent Young's modulus $E({\bf n})$ in BN and BP. The overall shape of the curve is similar for the two materials, with BN being less anisotropic than BP. Remarkably, BN is characterized by an exceptionally high Young's modulus, reaching 1916 GPa in the zigzag ($x$) direction, an order of magnitude larger than in BP. The least stiff direction corresponds to the direction of the layer stacking ($z$). Even in this case, the Young's modulus of compressed BN is found to be around 686 GPa vs 15 GPa in BP. The polycrystalline Young modulus calculated using Hill's averaging is found to be 1077 (57) GPa in BN (BP). The corresponding shear modulus amounts to 451 (23) GPa. The average Poisson's ratio is comparable in both materials, taking a value of 0.20.
\\

\section{\label{sec4}Conclusion}
Motivated by the recent synthesis of crystalline nitrogen with the orthorhombic A17 crystal structure, we have systematically studied electronic, optical, vibrational, and elastic properties of this compound at the experimental pressure conditions. To this end, we used density functional theory combined with the state-of-the-art quasiparticle self-consistent $GW$ approach with vertex corrections included in both the electronic and optical channels. Our analysis is focused on the comparison with black phosphorus, in order to elucidate mechanisms behind the difference between the two materials.

\begin{figure*}[t]
\includegraphics[scale=0.38]{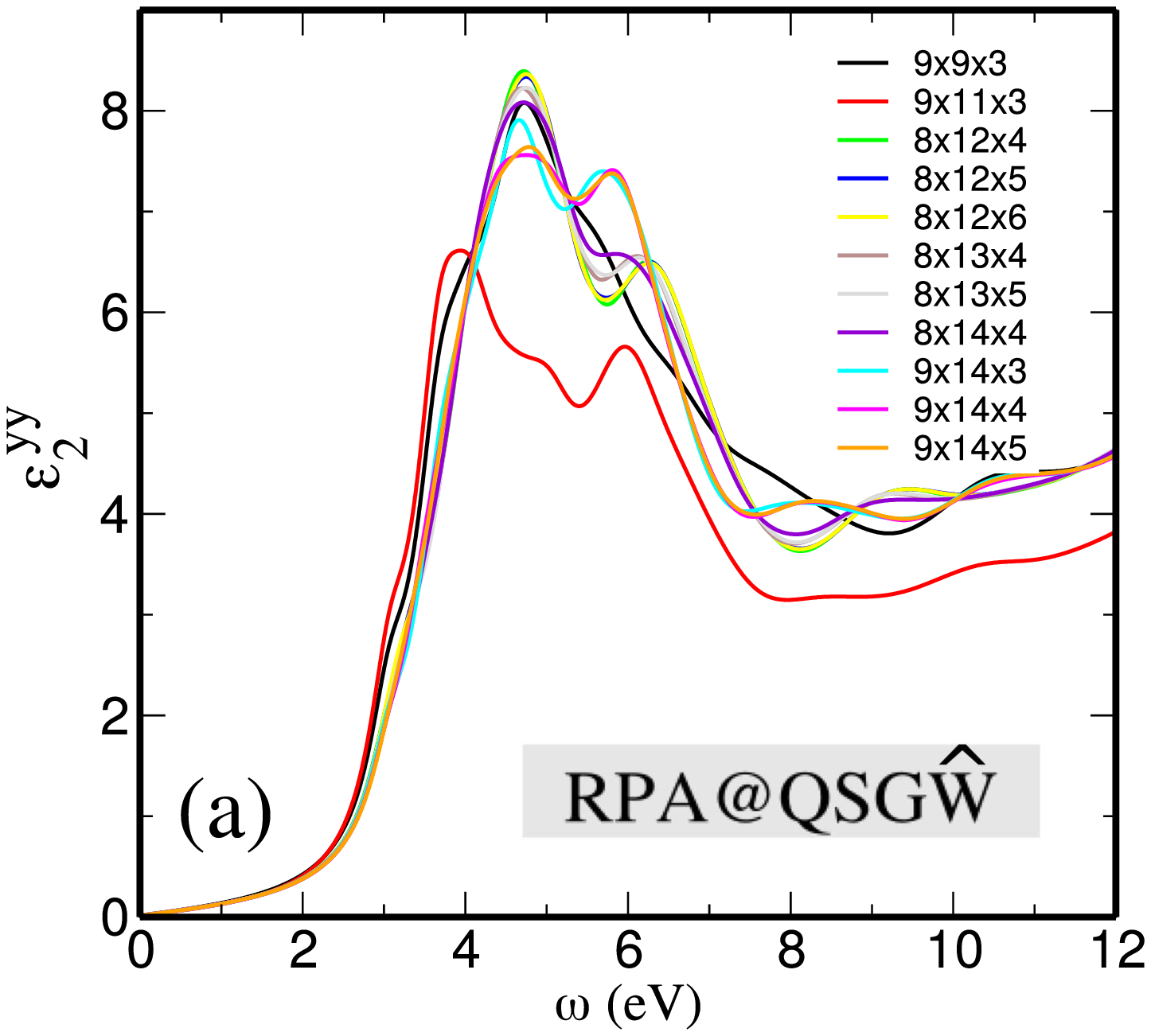}\,\,
\includegraphics[scale=0.38]{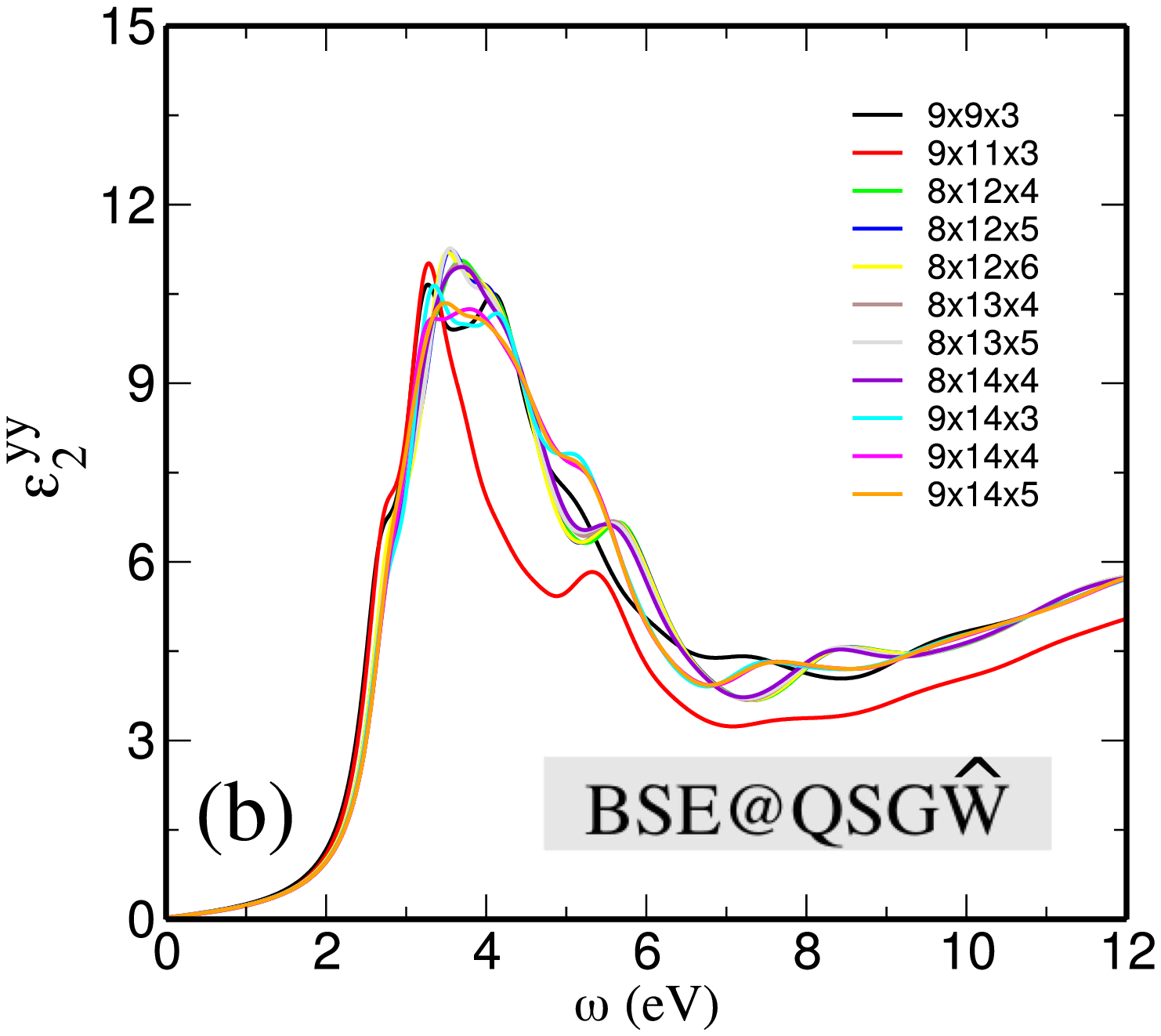}\quad
\includegraphics[scale=0.39]{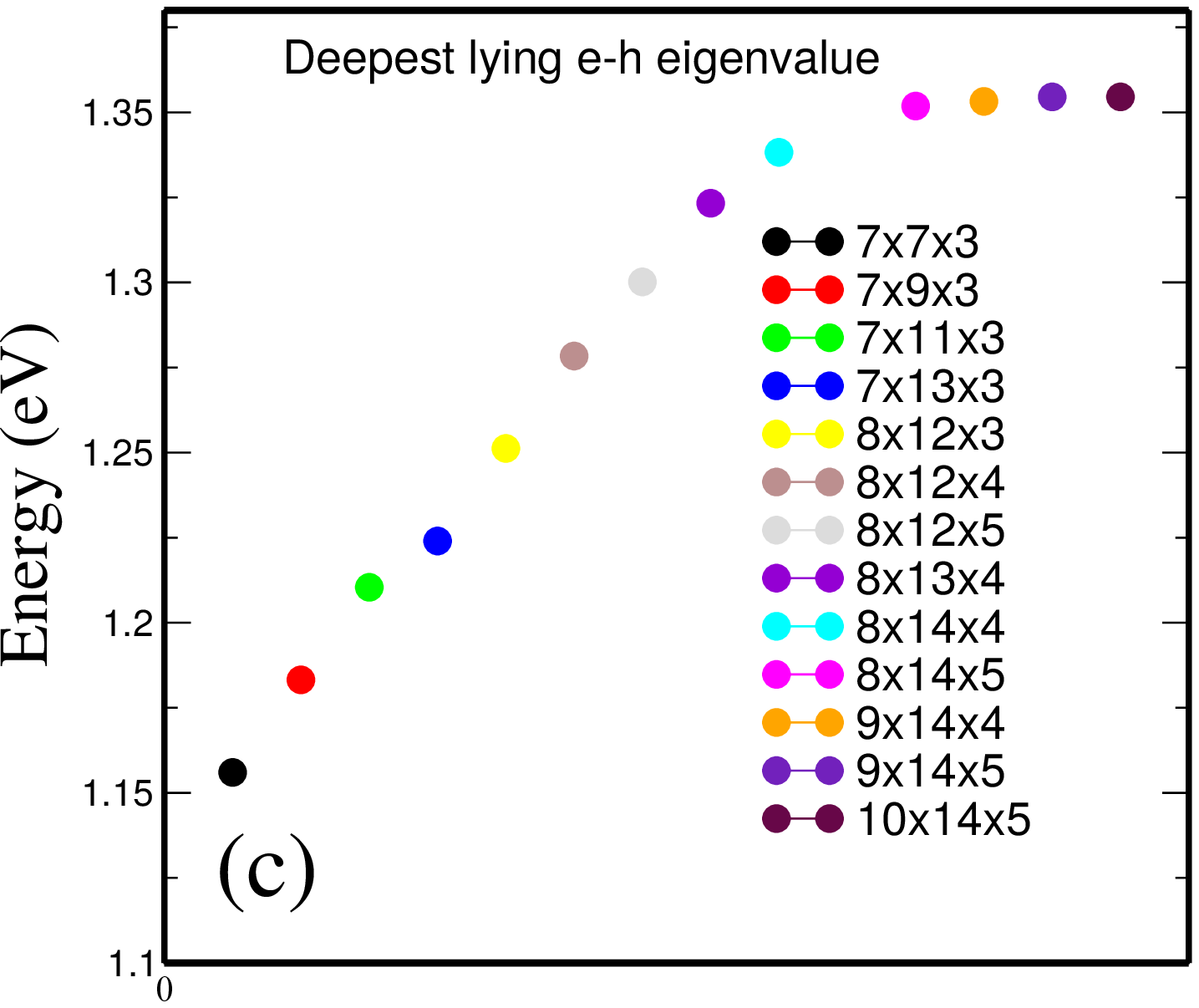}
\caption{\label{conv-test}
(a) and (b) The convergence of the RPA@QS$G\widehat{W}$ and BSE@QS$G\widehat{W}$ dielectric functions $\varepsilon_{2}^{yy}(\omega)$ in BN with respect to the {\bf k}-point mesh. (c) The convergence of the deepest-lying dark $e$-h eigenvalue.}
\end{figure*}

From the electronic and optical points of view, BN is characterized by a considerably larger optical gap (2.5 vs 0.26 eV), which ensures transparency in the visible spectral region. This difference can be ascribed to the large Coulomb interactions between the $p$ orbitals of nitrogen, resulting from a reduced dielectric screening. Despite the fact that the role of vertex corrections is significant in this case, it does not lead to the emergence of the excitonic peaks inside the gap, similar to BP. Unlike BP, the hole effective masses along the zigzag direction in BN are anomalously high ($m_x\sim 10m_e$), giving rise to a highly anisotropic electronic structure at the valence band edge.  

The spectrum of phonon excitations in BN spans a wider energy range compared to BP, which is not surprising. However, this difference cannot be solely explained by the atomic mass difference and indicates much stronger interatomic interactions in BN. This is further demonstrated by the exceptionally high stiffness of BN, which is comparable to that of diamond. Furthermore, BN is characterized by very high sound velocities, which are close to their ultimate limit. 
\\

\begin{acknowledgments}
The work of M.I.K., A.N.R., and S.A. was supported by the ERC Synergy Grant, Project No. 854843 FASTCORR.
F.T. and I.A.A. acknowledge support from the Knut and Alice Wallenberg Foundation (Wallenberg Scholar Grant No. KAW-2018.0194),
the Swedish Government Strategic Research Areas in Materials Science on Functional Materials at
Link\"oping University (Faculty Grant SFO-Mat-LiU No. 2009 00971) and  Swedish e-Science Research Center (SeRC), Swedish Research
Council (VR) Grant No. 2019-05600, and VINN Excellence Center Functional Nanoscale Materials (FunMat-2) Grant No. 201605156. 
A.V.P.'s calculations of the ground state properties of BN were carried out at the computer cluster at NUST ``MISIS'' and supported by RFBR, Project No. 20-02-00178.
M.v.S. and D.P. were supported by the U.S. Department of Energy, Office of Science, Basic Energy Sciences, Division of Chemical Sciences, under Contract No. DE-AC36-08GO28308.
The computations of phonons were enabled by resources
provided by the Swedish National Infrastructure for Computing (SNIC), partially funded by the
Swedish Research Council through Grant Agreement No. 2016-07213. S.A. acknowledges PRACE for awarding us access to Irene-Rome hosted by TGCC, France and Juwels Booster and Cluster, Germany. This work was also partly carried out on the Dutch national e-infrastructure with the support of the SURF Cooperative.
\end{acknowledgments}
\vspace{0.2cm}
\appendix*
\section{\label{appendix}$k$-point convergence of the optical absorption spectrum in BN}
Figure~\ref{conv-test} shows the convergence of the dielectric function $\varepsilon^{yy}_2(\omega)$ in BN with respect to the {\bf k}-point mesh calculated within RPA@QS$G\widehat{W}$ and BSE@QS$G\widehat{W}$. Additionally, the convergence of the deepest-lying dark $e$-h eigenvalue is also shown.


\bibliography{main}

\end{document}